\documentclass{article}
\usepackage{amssymb}
\usepackage{amsmath}
\usepackage{bm}
\usepackage{algorithm}
\usepackage[noEnd=true]{algpseudocodex}
\usepackage{xspace}
\usepackage{xparse}   % \NewDocumentCommand
\usepackage{xstring}
\usepackage{xifthen} % if-the-else see http://ctan.org/pkg/xifthen
\usepackage{boxedminipage}
\usepackage{graphicx}%%[draft] : do not embed figs/picts
\usepackage{float}
\usepackage{rotating}
\usepackage{url}
\usepackage{nicefrac}
\usepackage{comment}
\usepackage{soul}
\usepackage[normalem]{ulem} %% strikeout sout
\usepackage{verbatim}%%block comment
\usepackage{csquotes} % \textquote{}
\usepackage[T1]{fontenc}    % for accents
\usepackage[utf8]{inputenc} % for coding
\usepackage{pifont}
\usepackage{marvosym}
\usepackage{fullpage}

\usepackage[most]{tcolorbox}
\usepackage{hyperref}

\newcommand{\vmd}{\codecx{VMD}}
\newcommand{\pymol}{\codecx{PyMol}}
\newcommand{\tkinter}{\codecx{Tkinter}}
\newcommand{\panelhref}{\href{https://panel.holoviz.org/}{PANEL}}
\newcommand{\nglhref}{\href{https://nglviewer.org}{NGL}}

\NewDocumentCommand\latrans{m G{}}{{#1}^{\mathsf{#2 T}}}

\DeclareMathOperator*{\ArgminOp}{argmin}

\NewDocumentCommand{\argmin}{o}{%
  \IfNoValueTF{#1}%
    {\ArgminOp}%
    {\ArgminOp_{#1}}%
}

\NewDocumentCommand\sbulem {g}{\noindent $\bullet$\IfNoValueTF{#1}{}{{\em #1}}}
\NewDocumentCommand\sbullet{g}{\noindent $\bullet$\IfNoValueTF{#1}{}{{#1}}}
\newcommand{\sblwebhref}{\href{https://sbl.inria.fr}{Structural Bioinformatics Library}}

\newcommand{\codecx}[1]{{\tt \text{#1}}\xspace} %code convention
\newcommand{\ucainria}{\codecx{Universit\'e C\^ote d'Azur}, \codecx{Inria}, France\xspace}

\NewDocumentCommand\Sij{g}{\IfNoValueTF{#1}{\ensuremath{S_{ij}}}{\ensuremath{S_{#1}}}}
\NewDocumentCommand\Iij{g}{\IfNoValueTF{#1}{\ensuremath{I_{ij}}}{\ensuremath{I_{#1}}}}

\newcommand{\ie}{{\em i.e.}\xspace}
\newcommand{\eg}{{\em e.g.}\xspace}

\NewDocumentCommand\deltagd{s}{\IfBooleanTF{#1}{\Delta G_d^{\circ}}{\Delta G_d}}
\NewDocumentCommand\deltaga{s}{\IfBooleanTF{#1}{\Delta G_a^{\circ}}{\Delta G_a}}

\NewDocumentCommand\ddeltagd{s}{\IfBooleanTF{#1}{\deltagd*}{\deltagd}}

\NewDocumentCommand\dotpn{o m m}{%
\left<
#2, #3
\right>
\IfNoValueTF{#1}
{}
{_{#1}}
}

\newcommand{\calW}{\mathcal W}

\NewDocumentCommand\lrmsd{gg}{
\IfNoValueTF{#1}
{\text{lRMSD}\xspace}
{\text{lRMSD}\left(#1,#2\right)\xspace}
}

\NewDocumentCommand\rmsdcomb{gg}{%
\IfNoValueTF{#2}
{\ensuremath{\text{RMSD}_{\text{Comb.}}}\xspace}
{\ensuremath{\text{RMSD}_{\text{Comb.}}(#1,#2)}\xspace}
}

\definecolor{darkgreen}{rgb}{0,0.7,0}

\newcommand{\toblue}{\color{blue}\xspace}

\newcommand{\toblack}{\color{black}\xspace}

\newcommand{\makeremark}[2]{
  \newcommand{#1}[1]
    {$\longrightarrow$\textcolor{red}{\sc #2: ##1}$\leftarrow$ \medskip}}

\makeremark{\fc}{Frederic says}
\makeremark{\sm}{Simon says}

\definecolor{block-gray}{gray}{0.85}
\newtcolorbox{qcolgray}{colback=block-gray,boxrule=0pt,boxsep=0pt,breakable}

\NewDocumentCommand\dtms{O{k} g}{d^2_{#1}\IfNoValueTF{#2}{}{(#2)}}
\NewDocumentCommand\dtm{O{k} g}{d_{#1}\IfNoValueTF{#2}{}{(#2)}}
\NewDocumentCommand\dtmmed{g}{d_{DTMm}\IfNoValueTF{#1}{}{(#1)}}
\NewDocumentCommand\dtmemd{g}{d_{EMD}\IfNoValueTF{#1}{}{(#1)}}

\NewDocumentCommand\dwassk{O{k} G{}}{d_{\calW_{#1}}\IfNoValueTF{#2}{}{(#2)}}
\NewDocumentCommand\dpdzero{g}{d_{PD0}\IfNoValueTF{#1}{}{(#1)}}
\NewDocumentCommand\dpdone{g}{d_{PD1}\IfNoValueTF{#1}{}{(#1)}}
\NewDocumentCommand\dpdtwo{g}{d_{PD2}\IfNoValueTF{#1}{}{(#1)}}

\NewDocumentCommand\spectraldom{g}{\IfNoValueTF{#1}{\codecx{SPECTRALDOM}}{\codecx{SPECTRALDOM/#1}}}

\NewDocumentCommand\varinf{O{} g}{
\text{VI}_{#1}
\IfNoValueTF{#2}{}{(#2)}
}

\NewDocumentCommand\stiffc{G{}}{ \gamma^\text{Cov}_{#1} }
\NewDocumentCommand\stiffnc{G{}}{ \gamma^\text{NCov}_{#1} }
\NewDocumentCommand\stiffhbond{G{}}{ \gamma^\text{HB}_{#1} }
\NewDocumentCommand\stiffSB{G{}}{  \gamma^\text{SB}_{#1} }

\NewDocumentCommand\aij{G{ij}} {a_{#1}}
\NewDocumentCommand\bij{G{ij}} {b_{#1}}
\NewDocumentCommand\Bij{G{ij}} {B_{#1}}
\NewDocumentCommand\cij{G{ij}} {c_{#1}}
\NewDocumentCommand\Cij{G{ij}} {C_{#1}}

\NewDocumentCommand\dij{G{ij}} {d_{#1}}
\NewDocumentCommand\dijs{G{ij}}{d_{#1}^2}
\NewDocumentCommand\Dij{G{ij}} {D_{#1}}
\NewDocumentCommand\fij{G{ij}}{f_{#1}}
\NewDocumentCommand\gij{G{ij}}{g_{#1}}
\NewDocumentCommand\hij{G{ij}}{h_{#1}}
\NewDocumentCommand\iij{G{ij}}{i_{#1}}

\NewDocumentCommand\mij{G{ij}}{m_{#1}}

\NewDocumentCommand\Sd{G{d}}{S^{#1}}
\NewDocumentCommand\wij{G{ij}}{w_{#1}}
\NewDocumentCommand\expL{O{p} m}{
\exp
\IfEqCase{#1}{
{p}{(}      % p for parenthesis
{P}{\big(}
{b}{[}      % b/B for brackets
{B}{\big[}
{c}{\{}     % c/C for braces/curly brackets
{C}{\big\{}
}[\PackageError{expL}{Undefined option to expL: #1}{}]
#2
\IfEqCase{#1}{
{p}{)}
{P}{\big)}
{b}{]}
{B}{\big]}
{c}{\{}
{C}{\big\{}
}[\PackageError{expL}{Undefined option to expL: #1}{}]
}
\NewDocumentCommand\expl{O{p} m}{\expL[#1]{#2}}

\NewDocumentCommand\expX{O{} g}{%
\IfNoValueTF{#2} 
{ {\mathbb{E}_{#1}} }
{ {\mathbb{E}_{#1}}\left[ {#2} \right] }
}

\NewDocumentCommand\vvnorm{o m}{%
\left\lVert
{#2}
\right\rVert
\IfNoValueTF{#1}
{}
{_{#1}}
}

\newcommand{\frederic}{Fr\'ed\'eric\xspace}

\NewDocumentCommand\Calpha{g}{\IfNoValueTF{#1}{C_{\alpha}}{C_{\alpha;#1}}}
\NewDocumentCommand\Cbeta{g}{\IfNoValueTF{#1}{C_{\beta}}{C_{\beta;#1}}}

\newtheorem{example}{\noindent Example}{}
\newtheorem{remark}{Remark}

\NewDocumentCommand\paragraphmini{O{4pt} m}{%
{\vspace{#1}
\noindent{\bf #2}
}}
\NewDocumentCommand\paramini{O{4pt} m}{%
{\vspace{#1}
\noindent{\bf #2}
}}

\newcommand{\sbl}{\codecx{SBL}}
\newcommand{\sbllong}{Structural Bioinformatics Library\xspace}

\newcommand{\sblguigen}{\texttt{sbl-gui-generator.py}\xspace}

\title{
A Unified, Cross-Platform Framework for Automatic GUI and Plugin
Generation in Structural Bioinformatics and Beyond}

\author{Sikao Guo and Edoardo Sarti and \frederic Cazals,\\
  \ucainria \thanks{Correspondence: Frederic.Cazals@inria.fr}}

\begin{document}
\maketitle

\begin{abstract}
We present a workflow and associated toolkit to automate the creation
of graphical user interfaces (GUI) for executables run from command
line interfaces (CLI).
The workflow consists of three phases, namely (Step 1) the plugin
design, (Step 2) the formal (platform independent) specification of
the GUI, and (Step 3) the plugin code generation for the targeted
platforms.
Our architecture is aligned with the \emph{Model--View--Presenter} (MVP) 
pattern: steps one and two build the Model and View descriptions, while 
step three implements the Presenter layer that binds inputs, invokes the CLI, 
and updates outputs.
Once Step one has been (manually) completed, steps two and three are fully automated.
%%
%%This workflow eases the generation of coherent GUI for various interfaces and platforms.
The decoupled MVP design and platform-specific
generator modules enable reuse of logic, portability across
ecosystems, and significant reductions in engineering effort for
complex interactive applications.

We primarily use our workflow to generate GUI in structural
bioinformatics for CLI executables from the \sbllong, targeting three
platforms, namely \vmd,  \pymol,  and Web servers.
%% : \vmd~\cite{humphrey1996vmd}, \pymol~\cite{delano2002pymol}, and Web servers.
%% In particular, we successfully generated plugins for nine applications from the
%% \sbl, across these three platforms, demonstrating the ease of iteration and redesign 
%% inherent to our approach. \toblack
%%
%% To illustrate its versatility, we also generate plugins for a geometry processing
%% application, based on the JavaScript 3D library three.js.

The workflow can be used as a guideline,
while its implementation  available in the package 
\href{https://sbl.inria.fr/doc/Plugin_manager-user-manual.html}{ \toblue \codecx{Plugin\_manager}}
from the \sbllong can be used to effectively generate plugins.
\end{abstract}

\noindent{\bf Acronyms:} CLI: command line interface; GUI: graphical user interface,
MVP: Model--View--Presenter.


\section{Introduction}
%%i%%%%%%%%%%%%%%%%%%%%%%%%%%%%%%%%%%%%%%%%%%%%%%%%%%%%%%%%%%%%%%%%%%%%%%%%%%%%%%%

\subsection{Rationale}
%%ii-%-%-%-%-%-%-%-%-%-%-%-%-%-%-%-%-%-%-%-%-%-%-%-%-%-%-%-%-%-%-%-%-%-%-%-%-%-%-%

Graphical user interfaces (GUI) play a crucial role in making command
line interface (CLI) accessible to a broader range of users,
especially in scientific domains where complex analyses often rely on
scripts and terminal
workflows~\cite{Myers2000past,Deelman2018future}. However, creating
consistent GUIs for multiple platforms (desktop applications,
molecular-visualization environments, web interfaces) is often
laborious, requiring separate development efforts for each
(executable, platform) pair.  Moreover, standard software engineering
best practices---modularity, consistency, and maintainability---are
often compromised when manually porting interfaces across disparate
platforms.
\medskip

In this work, we present a workflow and associated toolkit designed to automate the creation of GUI plugins for CLI. 
  More specifically, our workflow enforces the aforementioned principles
automatically:
\begin{itemize}
\item \textbf{Separation of concerns (MVP).}  The overall architecture
  follows the Model--View--Presenter (MVP), where the Model handles
  the application's states, the View part focuses on the
  visualization, and the Presenter acts as the orchestra conductor.
With this pattern, the  generated code maintains a rigid boundary between state management,
GUI rendering, and execution logic, ensuring that business logic
remains independent of the visualization platform.

\item \textbf{Single source of truth.} Rather than maintaining
  fragmented codebases, our workflow derives all GUIs from a single
formal  specification. Modifications to the CLI or layout are propagated
  instantly to all targets, eliminating redundancy and synchronization
  errors.

%% By automating the generation of GUIs for several platforms, our
%% toolkit avoids specific developments for pairs (executable, platform),
%% and promotes consistency in interface design.  

\item \textbf{Interface consistency.} Automated generation ensures two
  levels of consistency: (i) vertical consistency, where the same
  plugin offers an identical look-and-feel across different host
  platforms; and (ii) horizontal consistency, where plugins for
  different CLI applications share a uniform logic, significantly
  reducing the user learning curve.

\item \textbf{Maintainability and extensibility.} Development is
  centralized: modifying the design specification automatically
  updates all downstream implementations. Furthermore, the modular
  architecture facilitates the integration of new platforms simply by
  adding a corresponding code generator.

\item \textbf{Standardized project structure.} The workflow enforces a
  canonical directory organization (\emph{design},
  \emph{specification}, \emph{generated code}), simplifying
  collaboration and the on-boarding of new developers.
\end{itemize}

\subsection{A use case in bioinformatics and beyond}
\label{sec:plugins-sbl}
%%ii-%-%-%-%-%-%-%-%-%-%-%-%-%-%-%-%-%-%-%-%-%-%-%-%-%-%-%-%-%-%-%-%-%-%-%-%-%-%-%

\paragraph{Bioinformatics.} 
To see how our strategy resonates in practice, we 
use our framework to provide various plugins for CLI applications
from the \sbllong, a comprehensive library whose architecture encompasses
optimized low level generic algorithms, as well as a number of
specific applications in structural
bioinformatics--\cite{cazals2017structural} and \sblwebhref[blue].  
We primarily target \vmd and \pymol, which are reference molecular
modeling environments enabling the visualization of biomolecules --
for an example see Fig. \ref{fig:plugin-example-intervor} and
\href{https://sbl.inria.fr/doc/Space_filling_model_interface-user-manual.html}{\toblue
  Intervor}.  These plugins are developed with \tkinter and \emph{Qt},
which are specific GUI libraries. In practice, the \vmd plugins use a
Tk/Tkinter stack, the \pymol plugins use \emph{pymol.Qt} -- details
in \href{https://sbl.inria.fr/doc/Plugin_manager-user-manual.html}{
  \toblue \codecx{Plugin\_manager}} documentation.

We also use our framework to provide the same applications as 
web-based plugins, implemented using \panelhref and \nglhref, 
enabling a fully Python-driven development workflow coupled with GPU-accelerated 
molecular rendering in the browser. This combination demonstrates that complex 
scientific visualization pipelines-traditionally tied to desktop tools such as 
\vmd and \pymol can be generated automatically and deployed seamlessly to the 
web, without manual JavaScript development.

This massive deployment highlights the ability of
our framework to streamline the maintenance and evolution (redesign,
updates) of scientific GUIs.

\paragraph{Geometry processing.}
To emphasize the versatility of our approach, we also provide a
geometry-processing use case, generating an interactive viewer for
geometric primitives (edges, triangles) built on three.js.
This cross-domain example highlights the architectural
generality of our system: the same metadata specification and code
generation pipeline produces functional GUIs across disparate
scientific fields, programming languages, and visualization back
ends. In particular, the decoupled MVP design and platform-specific
generator modules enable reuse of logic, portability across
ecosystems, and significant reductions in engineering effort for
complex interactive applications.

%% We validate our approach by generating plugins for nine applications
%% from the \sbllong, a comprehensive software environment providing
%% reference tools solving various problems in computational structural
%% biology / molecular modeling--see \cite{cazals2017structural} and the
%% \sblwebhref[blue].  We make all plugins available \vmd, \pymol, and
%% web platforms, see \toblack
%% \href{https://sbl.inria.fr/doc/sbl-plugins-guide.html}{\toblue SBL
%%   Plugins Guide}.  

\subsection{Main design choices}
%%ii-%-%-%-%-%-%-%-%-%-%-%-%-%-%-%-%-%-%-%-%-%-%-%-%-%-%-%-%-%-%-%-%-%-%-%-%-%-%-%

\begin{figure}[htb]% or !htb or H
  \centerline{\includegraphics[width=.75\linewidth]{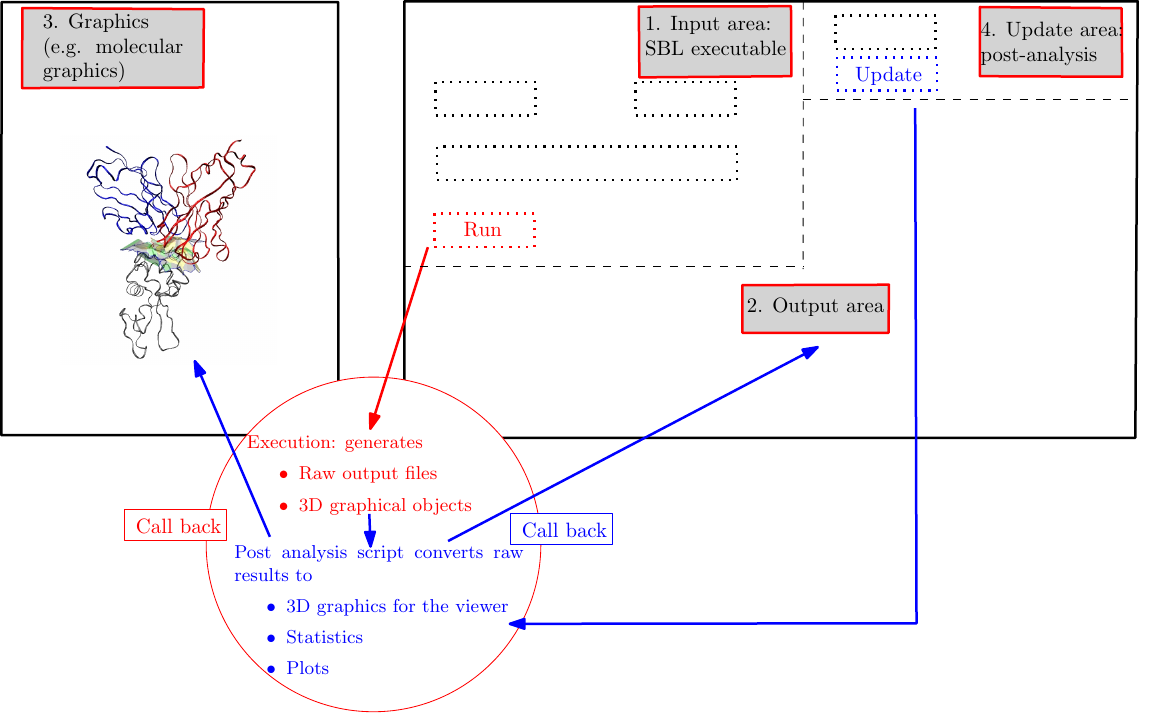}}
  \caption{{\bf GUI: generic architecture.}
  A plugin consists of four
    main areas: (1) the Input area, (2) The output area used to display statistics and figures,
    (3) The 3D graphics area, and (4) The optional update area to
    re-run post-analysis and refresh outputs and/or the 3D view.}
\label{fig:plugin-archi} 
\end{figure}

\paragraph{CLIs, platforms, and plugins.}
Given an executable run from the CLI, our toolkit automates the
generation of {\em plugins}, namely GUIs embedded in predefined {\em
  platforms}.
We assume that a plugin generically consists of three or four main areas
(Fig.~\ref{fig:plugin-archi}):
\begin{enumerate}
\item  An \emph{input area}, where the user
specifies options and selects input files passed to the CLI executable.

\item The \emph{output area} presenting 
artifacts (plots, images, and tables)
prepared by a callback call to a post-analysis  script 
 upon the termination of the CLI run.

\item The \emph{3D rendering engine} visualizing  structures, 
selections, and geometric overlays, also triggered by a callback function.

\item An optional \emph{update area} enabling 
additional controls to re-run post–analysis steps and selectively refresh 
the output area and/or the 3D display.
\end{enumerate}
A plugin also relies on a GUI library, for example \emph{Qt}, 
\emph{Tkinter}, and \emph{Panel}, and the choice of the GUI library
naturally depends on the targeted platform.

\begin{figure}[htb]% or !htb or H
	\centerline{\includegraphics[width=.5\linewidth]{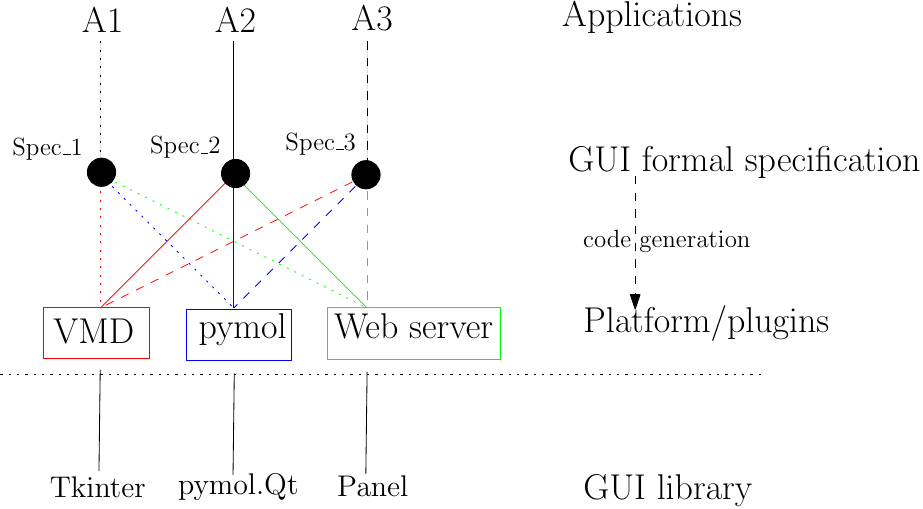}}
	\caption{{\bf Plugin generation using one specification file per application and one
			code generator per targeted platform.} 
		For each application $A_i$, one provides a GUI specification $S_i$.
		Each targeted platform is handled by a dedicated {\em code generator} $C_j$:
		$C_j(A_i)$ generates the code of i-th application for the j-th platform.
		Platforms are supported by various rendering engines.
	} 
	\label{fig:plugin-generation} 
\end{figure}

\paragraph{Design factoring: from a quadratic to a linear number of developments.}
As just noticed, a difficulty in developing plugins is to release
coherent versions for different platforms.  In our context, these
platforms include 3D molecular visualization applications such as
\href{https://www.ks.uiuc.edu/Research/vmd/}{\toblue
  \vmd}~\cite{humphrey1996vmd} and
\href{https://www.pymol.org/}{\toblue \pymol}~\cite{delano2002pymol},
a web environment using \href{https://nglviewer.org}{\toblue NGL
  Viewer}~\cite{rose2015ngl,rose2018ngl} for molecular visualization,
and the JavaScript 3D library \href{https://threejs.org/}{\toblue
  three.js} for geometric rendering and processing.  Our tool is
unique in bridging this gap, targeting both domain-specific desktop
hosts and web environments simultaneously. In theory, for $n$
applications and $m$ targets, $n\times m$ specific developments are
required.
We instead propose a solution requiring in $n+m$ developments using
(i) one formal specification per application, and (ii) code generator
modules which are platform specific
(Fig. \ref{fig:plugin-generation}). With this design, if the
underlying CLI changes (e.g., a new flag is added) or new output is
introduced, the developer updates a single specification file/UI
layout, and all supported platforms are updated instantly.

\paragraph{The Model–View–Presenter (MVP) framework.}
The automatically generated GUI is structured according to the 
{\bf Model–View–Presenter} (MVP) architecture~\cite{potel1996mvp}.
The {\bf Model} manages the application’s state and core logic, 
such as selected CLI options, file paths, execution status and post-analysis parameters.
The  {\bf View} part handles user interaction and renders the interface;
it 
defines widgets (labels, entries, buttons, output panels) and emits user events. It does
not contain any command-building or business logic.
The {\bf Presenter} finally mediates between them: it handles View
events, orchestrates CLI command construction and execution,
interprets results and updates both the Model and the View (including
3D viewer refreshes).

This separation of concerns keeps the user-interface layer light and focused 
on presentation, while the business logic and command-execution logic remain 
independent—thus enabling easier testing, maintenance and cross-platform 
portability across our supported environments (Tkinter/\vmd, Qt/\pymol, and Panel/Web).

\subsection{Availability}
%%ii-%-%-%-%-%-%-%-%-%-%-%-%-%-%-%-%-%-%-%-%-%-%-%-%-%-%-%-%-%-%-%-%-%-%-%-%-%-%-%

Our framework is documented in the package
\href{https://sbl.inria.fr/doc/Plugin_manager-user-manual.html}{
  \toblue \codecx{Plugin\_manager}} from the \sbllong.

Plugins of the \sbllong currently disseminated are documented on the
following
\href{https://sbl.inria.fr/doc/sbl-plugins-guide.html}{\toblue Plugins
  guide} page.  These plugins are also explained in short videos on
the following \href{https://www.youtube.com/@SBL-plugins}{\toblue
  @SBL-plugins channel}.

\section{Framework architecture and code generation}
%%ii-%-%-%-%-%-%-%-%-%-%-%-%-%-%-%-%-%-%-%-%-%-%-%-%-%-%-%-%-%-%-%-%-%-%-%-%-%-%-%

\subsection{Terminology and code generation overview}
%%ii-%-%-%-%-%-%-%-%-%-%-%-%-%-%-%-%-%-%-%-%-%-%-%-%-%-%-%-%-%-%-%-%-%-%-%-%-%-%-%

We use the following terminology:
\begin{itemize}
  \item \textbf{Frontend:} The set of widgets composing  the user interface, \ie
 labels, entries, checkboxes, buttons, and output panels. These widgets are 
  defined in a JSON specification and represent how the user interacts with the GUI.  
  \item \textbf{Backend:} The logic that converts frontend inputs into CLI commands, 
  executes the underlying executable, maps the results back to output widgets 
  (e.g., text, images, tables, pdfs, htmls), and updates control of the outputs. In other words, 
  it is the execution and result-handling layer of the GUI.  
  \item \textbf{GUI library:} The rendering engine used to instantiate
    the frontend widgets and connect them with backend
    logic. Supported libraries are \emph{pymol.Qt}, \emph{PyQt6},
    \emph{Tkinter}, and \emph{Panel}, allowing cross-platform GUI
    generation from the same JSON specification.
  \item \textbf{Platform:} The hosting environment in which the generated GUI runs. 
  This can be a \emph{\vmd plugin}, a \emph{\pymol plugin}, or a \emph{web application} 
  served from a local or remote server.  
\end{itemize}
\medskip

\begin{figure}[htb]% or !htb or H
\centerline{\includegraphics[width=.85\linewidth]{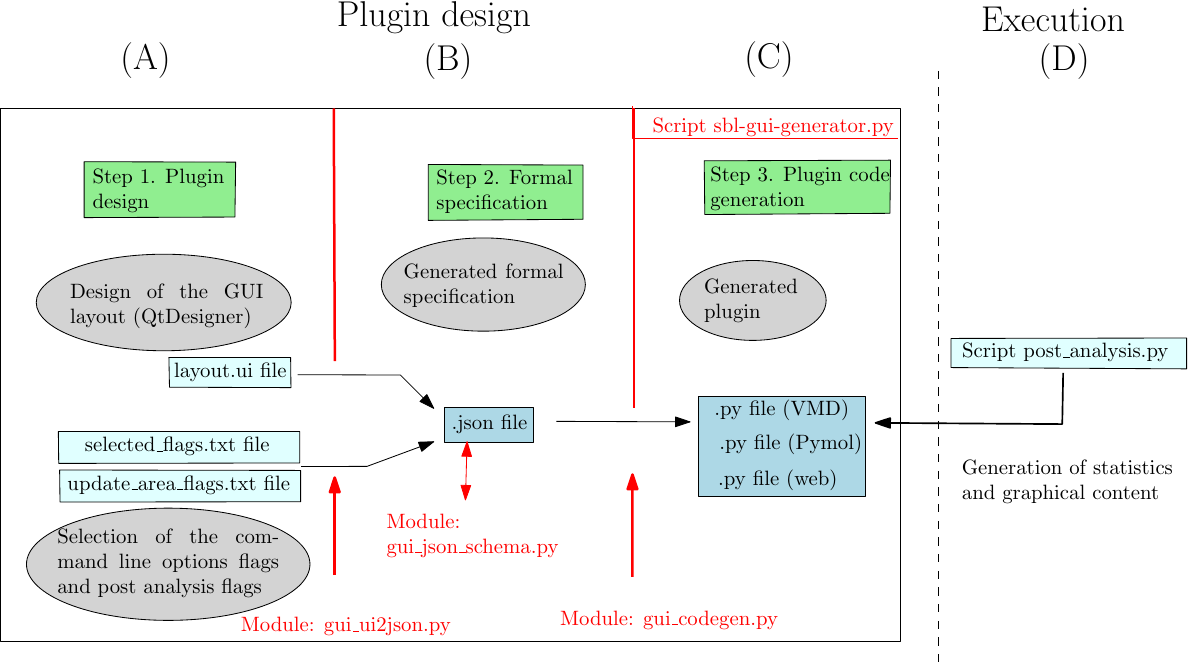}}
\caption{{\bf Plugin workflow generation with the three phases, and
    execution. Color conventions: light
    cyan: files prepared by the designer; light blue: files
    automatically generated; red: python modules provided in the
    package.}  {\bf (A)} Design phase. Using a layout editor, the
  designer (i) arranges the relative positions and sizes of widgets
  for the Input, Output, and/or Update areas; (ii) selects the CLI
  options (flags and arguments) to expose in the Input area; and (iii)
  specifies post–analysis parameters to expose in the optional Update
  area, whose changes selectively refresh the Output area and/or the
  3D display.
{\bf (B)} Generation of the formal specification.
A formal specification of the layout for the Input and Output areas is automatically generated.
{\bf (C)} Plugin generation. The python code of plugins is automatically generated.
{\bf (D)} Execution. The statistics and graphical contents are generated, and the 
GUI updated via callbacks.
} 
\label{fig:plugin-workflow} 
\end{figure}

Our  workflow involves three phases (Fig.~\ref{fig:plugin-workflow}):
\begin{enumerate}
\item (Step 1) Plugin design: choice of the options exposed, of the
  statistics/graphics displayed, and of the GUI layout--\ie arrangement of the
  input area, output areas, and/or update area.
\item (Step 2) Formal specification: automatic generation of a formal specification of the plugin using the JSON format.
\item (Step 3) Plugin code generation: automatic code generation for the targeted GUI library.
\end{enumerate}
In terms of the MVP architecture, Steps 1 and 2 build the Model and the View 
(i.e., the data/logic specification and the widget layout), while Step 3 
implements the Presenter (the generated code that binds the Model and View, 
handles user events, orchestrates CLI execution and updates). This clear 
separation supports modularity, maintainability and portability across multiple GUI platforms.

%% These steps are implemented by Python modules and one executable python
%% script (\sblguigen), each with a clearly separated responsibility
%% (Fig.~\ref{fig:plugin-workflow}). This design enables modularity,
%% reusability, and extensibility for integrating new GUI frameworks or
%% adapting different command-line applications. 

We now review these phases in detail.

\subsection{Step 1: plugin design}
\label{sec:step1}
%%ii-%-%-%-%-%-%-%-%-%-%-%-%-%-%-%-%-%-%-%-%-%-%-%-%-%-%-%-%-%-%-%-%-%-%-%-%-%-%-%

\begin{figure}[htb]
	\centering
	\includegraphics[width=0.95\textwidth]{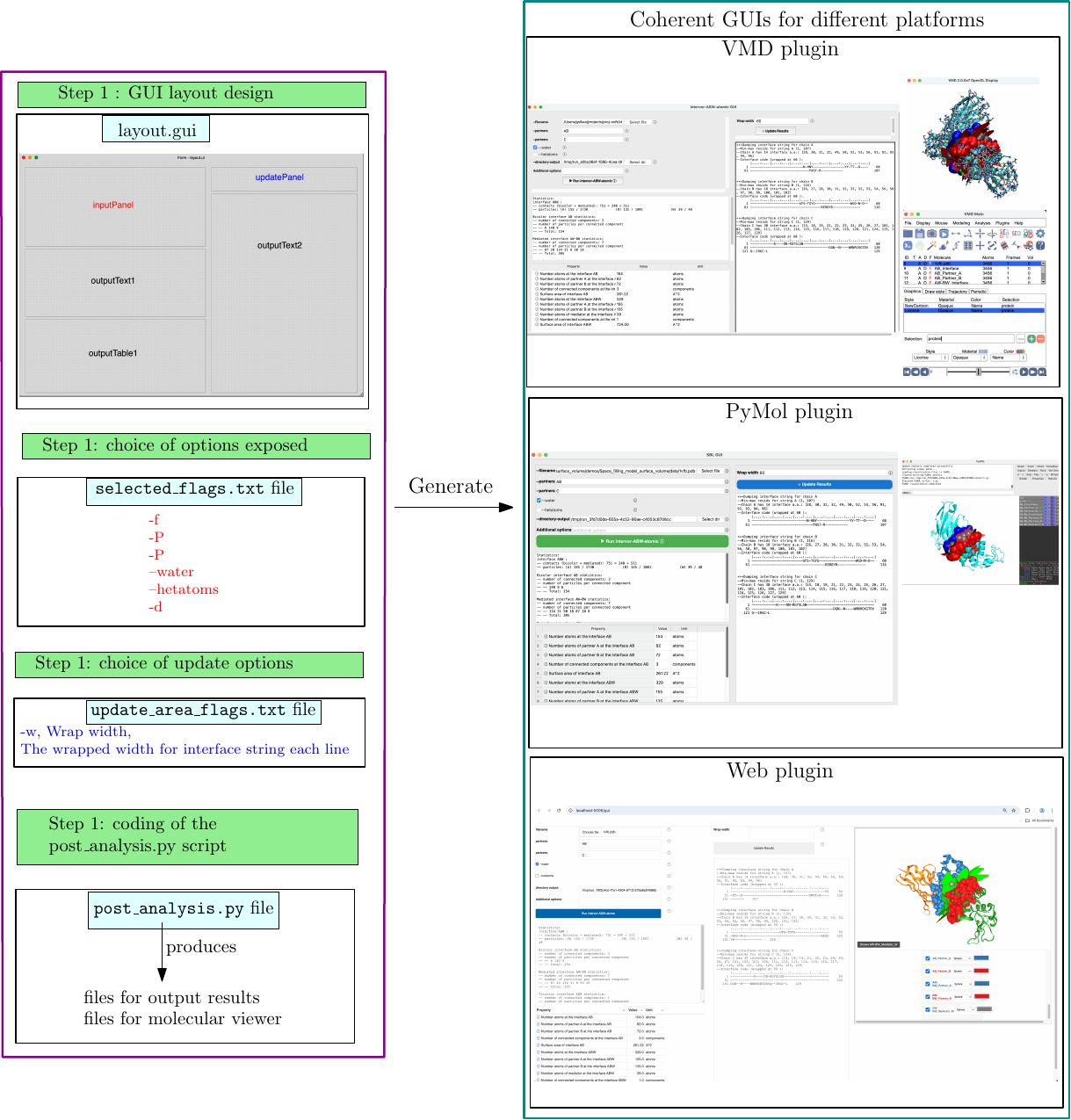}
	\caption{{\bf Plugin design illustrated for an application modeling the interface of a biomolecular complex.}
		Plugins for the CLI executable 
		\texttt{sbl-intervor-ABW-atomic.exe}
		from the SBL, see \cite{cazals2010modeling}.
		An example of using the \sblguigen to automatically create cross‑platform
		graphical user‑interface (GUI) plugins for command‑line applications.
		See also Fig. \ref{fig:plugin-workflow}.}
	\label{fig:plugin-example-intervor}
\end{figure}

This step uses two or three files (Fig. \ref{fig:plugin-example-intervor}):
\begin{enumerate}
\item A Qt Designer file defining the full layout for the Input and Output areas (named \texttt{layout.ui});
\item A plain–text list of CLI flags to expose (named \texttt{selected\_flags.txt});
\item (Optionally) A plain text list of post–analysis parameters exposed in the Update area (named \texttt{update\_area\_flags.txt}).
\end{enumerate}
The requirements for these files are detailed in the 
\href{https://sbl.inria.fr/doc/Plugin_manager-user-manual.html}{ \toblue \codecx{Plugin\_manager}} documentation.
\toblack

% \paragraph{Requirements for the widgets in the \texttt{layout.ui} file.}
% The requirements are as follows:
% \begin{enumerate}
% \item Each widget (input area, update area, or output areas) stores a \texttt{geometry} object with \texttt{x}, \texttt{y}, \texttt{width}, \texttt{height}.
% \item The Input area widget is named \texttt{inputPanel}. The optional Update area widget is named \texttt{updatePanel}.
% \item Output widgets are named \texttt{outputText\textbackslash d+}, \texttt{outputFigure\textbackslash d+}, \texttt{outputPDF\textbackslash d+}, 
% \texttt{outputHtml\textbackslash d+}, \newline
% \texttt{outputHtmlTabs}, or \texttt{outputTable\textbackslash d+} according to their type.
% \end{enumerate}

% Consequently, to ensure consistency with widget names, the naming
% conventions for the files produced by the post-analysis script are as
% follows:
% %%
% \texttt{outputText1} expects a text file (e.g., \texttt{outputText1.txt});
% \texttt{outputFigure1} expects \texttt{outputFigure1.png} for images
% or \texttt{outputFigure1.dat} for the raw data plotted in interactive figures in GUI; 
% \texttt{outputPDF1} expects a PDF file (e.g., \texttt{outputPDF1.pdf});
% \texttt{outputHtml1} expects an HTML file (e.g., \texttt{outputHtml1.html});
% \texttt{outputHtmlTabs} expects multiple HTML files;
% \texttt{outputTable1} expects a CSV file (e.g., \texttt{outputTable1.csv}).
% \toblack

\subsection{Step 2: formal specification}
\label{sec:step2}
%%ii-%-%-%-%-%-%-%-%-%-%-%-%-%-%-%-%-%-%-%-%-%-%-%-%-%-%-%-%-%-%-%-%-%-%-%-%-%-%-%

The \sblguigen script first converts the layout and flags files
(\texttt{layout.ui}, \texttt{selected\_flags.txt},
\texttt{update\_area\_flags.txt}(optional)) to a JSON specification
that records widget types, geometry, and metadata.  The JSON
specification is a structured representation of the GUI design,
capturing the layout and behavior of widgets in a platform-agnostic
format.  The detailed description of the JSON specification is
provided in the \codecx{Plugin\_manager} documentation.

Step two relies on the following two modules:
\begin{enumerate}
\item The module \texttt{gui\_json\_schema.py} defines a JSON schema, dataclasses, and 
helper functions to parse and validate GUI specifications. This module is used to
define the formal specification.

\item  The module  \texttt{gui\_ui2json.py} converts Qt Designer \texttt{layout.ui} files into JSON skeletons,
and merges command-line tool flags into a structured input panel, and optional update-area flags into
a structured update panel. 
\end{enumerate}

\subsection{Step 3: plugin code generation}
%%ii-%-%-%-%-%-%-%-%-%-%-%-%-%-%-%-%-%-%-%-%-%-%-%-%-%-%-%-%-%-%-%-%-%-%-%-%-%-%-%

Starting from the JSON GUI description, we automatically synthesize
application code organized according to the
\emph{Model–View–Presenter} (MVP) pattern for the target toolkit
(pymol.Qt, PyQt6, Tkinter, or Panel): Widget specifications are mapped
to toolkit-specific implementations in the View layer (See
specifications in the online user manual at
\href{https://sbl.inria.fr/doc/Plugin_manager-user-manual.html}{
  \toblue \codecx{Plugin\_manager}}).

The Presenter and Model layers—shared across all plugins—build and execute the CLI command, 
initiate optional post-analysis, and propagate updates to output widgets and the associated 3D molecular viewer.

% Three modules are responsible for the generation of the GUI specific
% code: \texttt{gui\_codegen\_pymol.py}, \toblack \texttt{gui\_codegen\_PyQt6.py},
% \texttt{gui\_codegen\_Tkinter.py}, and
% \texttt{gui\_codegen\_panel.py}.

%% The module \texttt{gui\_codegen.py} generates runnable source code for the selected
%% target framework (PyQt6, Tkinter, or Panel) based on the validated JSON layout. It delegates
%% the framework–specific code generation to \texttt{gui\_codegen\_PyQt6.py},
%% \texttt{gui\_codegen\_Tkinter.py}, and \texttt{gui\_codegen\_panel.py}.

% \begin{remark}
% \textbf{Web visualization (NGL Viewer / three.js).}
% For Panel targets, the generated interface embeds an NGL Viewer iframe for molecular structures 
% or a three.js iframe for geometric primitives. The iframe is reloaded after each execution to 
% present the updated molecular model or geometry output.
% \end{remark}

\subsection{Application specific generated files}
%%ii-%-%-%-%-%-%-%-%-%-%-%-%-%-%-%-%-%-%-%-%-%-%-%-%-%-%-%-%-%-%-%-%-%-%-%-%-%-%-%

As explained in Sec. \ref{sec:plugins-sbl}, we develop plugins for the \sbllong.
For each application, say \texttt{\$SBL\_DIR/Application\_XXX},
the \texttt{plugins/} directory contains exactly three subfolders in
addition to the driver script:
\begin{itemize}
    \item \texttt{step1\_design/}: inputs for the generator, including \texttt{layout.ui}, \texttt{selected\_flags.txt}, and optional \texttt{update\_area\_flags.txt}.
    \item \texttt{step2\_formal\_spec/}: the auto–generated formal specification in JSON format.
    \item \texttt{step3\_generated\_code/}: the auto–generated plugin code, packaged as \texttt{vmd.tar.gz}, \texttt{pymol.tar.gz}, and \texttt{web.tar.gz}.
\end{itemize}

\subsection{Post-analysis}
%%ii-%-%-%-%-%-%-%-%-%-%-%-%-%-%-%-%-%-%-%-%-%-%-%-%-%-%-%-%-%-%-%-%-%-%-%-%-%-%-%

As already noticed (Fig. \ref{fig:plugin-archi}), a plugin also
requires a script parsing the raw results generated by the executable,
so as to convert the outputs to the required figures, tables, or texts
for the GUI Output areas, and molecular / geometric viewer files for display.
We assume this task is accomplished by the
script \texttt{post\_analysis.py}, which is platform independent.

\section{Examples}
%%i%%%%%%%%%%%%%%%%%%%%%%%%%%%%%%%%%%%%%%%%%%%%%%%%%%%%%%%%%%%%%%%%%%%%%%%%%%%%%%%

\subsection{Example 1: structural bioinformatics}
%%ii-%-%-%-%-%-%-%-%-%-%-%-%-%-%-%-%-%-%-%-%-%-%-%-%-%-%-%-%-%-%-%-%-%-%-%-%-%-%-%

% \fc{I liked the fact to have one command line showing how simple it is to generate the plugins. may be put it back into a remark env?}
\paragraph{Application.} We show how to generate plugins for the
the CLI executable is \texttt{sbl-intervor-ABW-atomic.exe} from the
\sbl, which implements a model for biomolecular interfaces based on
Vornoi/power diagrams (Fig. \ref{fig:plugin-example-intervor},
\cite{cazals2010modeling} and
\href{https://sbl.inria.fr/doc/Space_filling_model_interface-user-manual.html}{\toblue Intervor}).

\paragraph{Generation.}
Once the user defined files have been designed, the generation of plugins
is straightforward (Example \ref{ex:gen-command})
The \texttt{----format} option selects the GUI backend for code generation
(Fig.~\ref{fig:plugin-generation}):
\begin{itemize}
\item \texttt{pyqt} generates a \pymol{} plugin (pymol.Qt),
\toblack
\item \texttt{tkinter} emits a \vmd{} plugin (Tk/Tkinter),
\item \texttt{panel--ngljs} builds a Panel + NGL Viewer web application.
\end{itemize}
The companion \texttt{----gui-output} option specifies the directory in which 
the generated source code is written, with a sensible default that depends on the selected backend.
The \texttt{----format} option accepts multiple values, allowing users to produce several GUI backends 
in a single command.

Across all targets, the same \texttt{layout.ui}, \texttt{selected\_flags.txt}, \texttt{update\_area\_flags.txt}, 
and post-analysis script are reused; only the generated GUI code and the corresponding loading mechanism 
differ between environments.

\paragraph{Usage.}
For convenience, we provide a wrapper script
\texttt{run\_generator.sh}
in\\ 
\texttt{\$SBL\_DIR/Applications/Space\_filling\_model\_interface/plugins/},
which automatically runs the appropriate code-generation pipeline for
\texttt{sbl-intervor-ABW-atomic.exe}
(Fig. \ref{fig:plugin-example-intervor}.  The script generates all
three GUI backends, places them in their respective output
directories, and prepares the auxiliary files required by each
platform.

\begin{example}
\label{ex:gen-command}
  \textbf{Plugin generation command.}  
  The command below creates the three GUI back-ends for the executable \texttt{sbl-intervor-ABW-atomic.exe}:

\begin{verbatim}
sbl-gui-generator.py --ui layout.ui --exe sbl-intervor-ABW-atomic.exe \
    --flags selected_flags.txt --post-script intervor_abw_atomic_plugin_post_analysis.py \
    --update-flags update_area_flags.txt --format pyqt tkinter panel-ngljs\
    --gui-output generated_plugins/
\end{verbatim}
After execution, the directory \texttt{generated\_plugins/} contains separate native GUI implementations 
for VMD, PyMOL and Web platforms.
\end{example}

\subsection{Example 2: geometry processing}
%%ii-%-%-%-%-%-%-%-%-%-%-%-%-%-%-%-%-%-%-%-%-%-%-%-%-%-%-%-%-%-%-%-%-%-%-%-%-%-%-%

As an elementary geometry processing application we show how to
generate plugins displaying simplices of the celebrated
$\alpha$-complex, a simplicial complex subset of the Delaunay
triangulation-- \cite{edelsbrunner1992weighted} and
\href{https://doc.cgal.org/latest/Alpha_shapes_3/group__PkgAlphaShapes3Ref.html}{\toblue CGAL 3D alpha shapes}.
This construction is also relevant in structural bioinformatics since
selected simplices of the $\alpha$-complex code the combinatorial
structure of the boundary of a union of balls -- say atoms.

\paragraph{Generation.}
For \texttt{----format} option, there is another supported target:
\begin{itemize}
\item \texttt{panel--threejs} builds a Panel + Three.js web application.
\end{itemize}

\paragraph{Usage.}
For convenience, we provide a wrapper script \texttt{run\_generator.sh} in\\
\texttt{\$SBL\_DIR/Core/Alpha\_complex\_of\_molecular\_model/plugins/}, 
which automatically runs the appropriate code-generation pipeline for \texttt{sbl-alpha-complex-of-molecular-model.exe}.
The script generates the Panel + Three.js web application.

\begin{remark}
To generate a standalone application in addition to the web app, one
may proceed as follows.  Assume that the CLI generates a 3D output,
\eg .ply or .obj or .py file.  The post-analysis script processes it,
generating a .py file visualized by say the
\href{https://vedo.embl.es/}{\toblue V$\ni$do} viewer in the output
area.
\end{remark}

\begin{comment}
\begin{itemize}
\item Singular: the simplex $\sigma$ does not have any coface
\item Interior: the simplex $\sigma$ has all its cofaces from the Delaunay triangulation
\item Regular otherwise.
\end{itemize}

This app offers the possible options:
* choice of alpha
* choice of the type of sipmlex displayed: edges, triangles, cells

* choice of the tag for simplices: singular, regular, interior. (post-analysis)
./sbl-alpha-complex-of-molecular-model.exe -f 1ycr.pdb -e -t -c --alpha 10
\end{comment}

\subsection{Specific features for different platforms and applications}
%%ii-%-%-%-%-%-%-%-%-%-%-%-%-%-%-%-%-%-%-%-%-%-%-%-%-%-%-%-%-%-%-%-%-%-%-%-%-%-%-%

\paragraph{Integration with molecular graphics environments.} 
The GUIs support seamless communication with three widely used molecular
graphics platforms: \vmd, \pymol, and Panel (via NGL Viewer / Three.js). 
The integration design distinguishes between (i) generic communication 
mechanisms (e.g., sockets, APIs, or file exchange) and (ii) specific 
functionalities for molecular and geometric visualization.  
\medskip

For \vmd, communication relies on a lightweight socket server. The GUI script
opens a TCP connection to the port specified by \texttt{VMDSOCK} (default 5555)
and sends visualization commands (e.g., \texttt{vmd\_visualize\_sbl\_plugin <post\_analysis\_output\_dir>}).
On the VMD side, a Tcl socket server listens, then executes the VMD commands
defined in the \texttt{.vmd} scripts produced by the post–analysis script.

\medskip
For \pymol, communication is performed via the \texttt{pymol.cmd} API.
The GUI loads the \texttt{.py} command files generated by the post–analysis script (e.g., using \texttt{cmd.do()})
to reproduce predefined molecular selections and representations inside PyMOL.

\medskip
For the \textbf{Panel web plugin}, communication uses an embedded NGL viewer for molecular modeling plugins
or Three.js for general 3D visualization.
The post–analysis script generates PDB and JSON descriptors (structures, selections, geometry);
the Panel app reloads the NGL / Three.js iframe (with cache‑busting) so the viewer reconstructs the 3D scene.

\medskip
This architecture yields consistent structural and surface visualization across
all environments while adapting to each platform’s communication mechanism
and rendering capabilities. It is easy to extend to support additional visualization platforms
by implementing the corresponding communication layer and viewer integration.

\section{Outlook}
%%i%%%%%%%%%%%%%%%%%%%%%%%%%%%%%%%%%%%%%%%%%%%%%%%%%%%%%%%%%%%%%%%%%%%%%%%%%%%%%%%

Looking forward, we envision several directions to further enhance the capabilities of our framework and streamline the development of scientific GUI.

\paragraph{AI-assisted GUI design.}
While our current workflow requires a layout and a flag pre-selection,
large language models (LLMs)~\cite{Brown2020NEURIPS} may be used to
assist in the design phase. A second avenue is automating the
preparation of the post-analysis scripts that convert raw outputs into
structured GUI artifacts. In the future, we anticipate program
synthesis mechanisms - again driven by LLMs - to infer post-processing
steps from examplar outputs or informal task descriptions. Such a
capability would elevate our pipeline from automated GUI generation to
automated analysis-UI co-design, reducing developer burden while
improving reproducibility.

\paragraph{Support for additional platforms and 3D engines}
Our current code generators target three environments - Tkinter / \vmd, pymol.Qt  / \pymol, and Panel / NGL Viewer or Three.js. The decoupled architecture opens the door to new front-ends and visualization engines, including but not limited to: Mol* ~\cite{sehnal2021molstar} for advanced web-scale molecular visualization; Unity or Unreal Engine for VR/AR scientific visualization~\cite{p2025ascribexr}; Blender for geometry processing or simulation animation; Jupyter-based widgets.

\paragraph{Toward autonomous scientific assistants}
Ultimately, the combination of AI-augmented specification, platfrom-independent GUI generation, and automated analysis scripts suggests a broader paradigm: an autonomous assistants for scientific UI construction, capable of taking a CLI tool, understanding its interface and output, and producing interactive applications across multiple enviroments with minimal human guidance. Such assitants would accelerate dissemination of scientific software, reduce onboarding friction, and strengthen reproducibility standards.

\medskip

\noindent{\bf Acknowledgments.}
This work is supported by a grant from the French government, managed by the French National Research Agency (ANR) under the France 2030 program, reference ANR-24-RRII-0002, and operated by the Inria Quadrant Programme (PIQ).

\bibliographystyle{unsrt}
%\bibliography{\wmybib/biogeom,\wmybib/mcs,\wmybib/abs,\wmybib/abs-sub}
\bibliography{local.bib}
%\clearpage
%\input{plugins-SI.tex}

%\clearpage
%\tableofcontents

%%

%% For journals like Proteins etc
%\renewcommand{\figurename}{Figure S\hspace{-.05cm}~}
%\renewcommand{\tablename}{Table S~}
%\setcounter{figure}{0}
%\setcounter{table}{0}
%\makeatletter %% package algorithm
%\renewcommand{\ALG@name}{Algorithm S\hspace{-.1cm}~}
%\makeatother
\end{document}